\documentclass[preprint2]{aastex}

\slugcomment{To be submitted to \emph{Astrophysical Journal}}

\begin{document}

\title{On the Simultaneous Generation of High-energy Emission and Submillimeter/Infrared
Radiation from Active Galactic Nuclei}

\author{Osmanov Z.\altaffilmark{1}}
\affil{Centre for Theoretical Astrophysics, ITP, Ilia State
University, Kazbegi str. 2a, 0160 Tbilisi, Georgia}



\altaffiltext{1}{z.osmanov@iliauni.edu.ge}

\begin{abstract}
For active galactic nuclei (AGNs) we study the role of the mechanism
of quasi-linear diffusion (QLD) in producing the high energy
emission in the MeV-GeV domains strongly connected with the
submillimeter/infrared radiation. Considering the kinetic equation
governing the stationary regime of the QLD we investigate the
feedback of the diffusion on electrons. We show that this process
leads to the distribution of particles by the pitch angles, implying
that the synchrotron mechanism is no longer prevented by energy
losses. Examining a reasonable interval of physical parameters, we
show that it is possible to produce MeV-GeV $\gamma$-rays, strongly
correlated with submillimeter/infrared bands.
\end{abstract}

\keywords{galaxies: active --- gamma rays --- submillimeter
--- infrared radiation}

\section{Introduction}

According to the model of AGNs, cold material close to the central
black hole forms an accretion disc, matter inside which due to the
dissipative forces transports inwards causing the accretion disc to
heat up. Such a hot material in turn can inverse-Compton scatter
photons up to X-ray energies \citep{blan}. From high energy
astronomical sources a special interest deserve blazar type AGNs the
standard model of which implies the presence of the supermassive
black hole, surrounded by the accretion disc and ejecting twin
relativistic jets. The observationally evident broadband emission
spectrum of blazars is made of two components: the low energy (from
radio to optical) domain attributed to the synchrotron emission and
the high energy (from $X$-rays to $\gamma$-rays) part formed by
either the inverse-Compton mechanism \citep{blan} or the curvature
radiation \citep{g96,tg05}. A recent investigation of the parsec
scale jets is very important. \cite{giro} argued that the high
energy and radio emissions are strongly correlated. Our model on the
other hand, as we will see, automatically provides a connection of
radiation in high energy and radio domains. Magnetospheres of AGNs
have strong magnetic fields, therefore, synchrotron cooling
timescales are relatively short, leading to efficient energy losses.
This in turn creates appropriate conditions for particles to transit
to their ground Landau state. When this happens, relativistic
electrons will move only along magnetic field lines without emitting
in the synchrotron regime.

Despite the very strong magnetic field, which prevents a continuous
process of the synchrotron emission, there is a possibility to
overcome the dissipative factors and maintain the radiation
mechanism. \cite{machus1} have studied the cyclotron instability of
two-component electron-positron plasma for the pulsar, NP 0532. It
was found that the instability arises near the light cylinder
surface (a hypothetical zone, where the linear velocity of rigid
rotation equals exactly the speed of light) leading to a certain
distribution of particles by pitch angles and the consequent
synchrotron radiation. \cite{lomin} considered the magnetospheres of
the pulsar NP 0532 and the Crab nebula, studying the generation of
waves from optical to gamma-ray domains. The similar approach was
presented by \cite{malmach} where the QLD was applied to the radio
pulsars. The authors found that the transverse momenta of
relativistic particles induced by the cyclotron instability caused
the stable non-zero pitch angle distribution maintained by means of
the QLD. Analyzing the data obtained from MAGIC Cherenkov telescope
between 2007 October and 2008 February \citep{magic} we found that,
the observed coincidence of signals in the optical and $\gamma$-ray
domains are easily explained by the QLD process, which leads to the
increase of the pitch angles, making the synchrotron process
feasible \citep{difus,difus1}.

In the magnetospheres of AGNs the magnetic fields are of the order
of $10^4G$ \citep{paradig}, close to the supermassive black hole,
and $100G-300G$ close to the light cylinder surface. Therefore, the
aforementioned QLD mechanism could be of great importance for AGNs
as well. For this purpose by considering the cyclotron instability
excited in the radio domain, in \citep{difus3} we studied the
quasi-linear interaction of proper modes of AGN magnetospheric
plasmas with the resonant plasma particles investigating the QLD in
the context of producing the soft and hard X-ray emission from AGNs.
Under favorable conditions this mechanism could also be efficient
for explaining the MeV-GeV energy synchrotron emission, strongly
connected either with the submilimeter radio band, or with the
infrared emission induced by the cyclotron instability. This will be
the subject of the present paper, which is organized as follows. In
Section 2 we describe our model, in Sect. 3 we apply the mechanism
of QLD to AGNs and in Sect. 4 we summarize our results.

\section{Main consideration}

In general, AGN magnetospheres consist of relatively low energy
particles and very high energy particles (electrons). Therefore, in
the framework of the model we consider the plasma composed of two
components: a) the so-called plasma component with the Lorentz
factor, $\gamma_p$ and b) the beam component with the Lorentz
factor, $\gamma_b$ ($\gamma_b\gg\gamma_p$). Such a system as was
shown by \citep{kmm} undergoes the cyclotron instability induced by
the Doppler effect with the following resonance condition
\begin{equation}\label{cycl}
\omega - k_{_{\|}}V_{_{\|}}-k_xu_x\pm\frac{\omega_B}{\gamma_b} = 0,
\end{equation}
where $k_{_{\|}}$ is the longitudinal (parallel to the background
magnetic field) component of the wave vector, $k_x$ is the component
along the drift, $V_{_{\|}}$ is the longitudinal component of plasma
flow velocity, $u_x\equiv cV_{_{_{\|}}}\gamma_b/\rho\omega_B$ is the
drift velocity of particles, $c$ is the speed of light, $\rho$ is
field line's curvature radius, $\omega_B\equiv eB/mc$ is the
cyclotron frequency, $B$ is the magnetic induction and $e$ and $m$
are electron's charge and the rest mass, respectively. Positive sign
corresponds to the damping of the excited modes, whereas the
negative sign relates to the unstable mode. One can show that, when
the aforementioned resonance takes place, the transverse waves with
the dispersion relation
\begin{equation}\label{disp1}
\omega_t \approx kc\left(1-\delta\right),\;\;\;\;\;\delta =
\frac{\omega_p^2}{4\omega_B^2\gamma_p^3},
\end{equation}
are induced. $k$ is the modulus of the wave vector, $\omega_p \equiv
\sqrt{4\pi n_pe^2/m}$ is the plasma frequency and $n_p$ is the
plasma density. From Eqs. (\ref{cycl},\ref{disp1}) is clear that the
excited cyclotron frequency is given by \citep{malmach}
\begin{equation}\label{om1}
\omega\approx \frac{\omega_B}{\delta\cdot\gamma_b}.
\end{equation}
In spite of the resonant character of the cyclotron modes, the
corresponding frequency is not well peeked, because $\omega$ depends
also on the Lorentz factors of the resonant (beam) particles, that
do not have narrow energy spectra. It is worth noting that unlike
the synchrotron mechanism ($\lambda<n_p^{-1/3}$), ($\lambda$ is the
wavelength), where radiation process can be described by a single
particle approach, excitation of the aforementioned waves is a
collective phenomenon ($\lambda>n_p^{-1/3}$), which in its turn is a
direct consequence of one-dimensionality of the distribution
function. On the other hand, such a behaviour of this function is
guaranteed by the strong magnetic field that forces particles to
move along the field lines.

By the first sentence after Eq. 3 we would like to note that despite
the resonant character of the process, the corresponding frequency
is not well peeked, because the frequency depends also on the
Lorentz factor of the resonant particles, and the corresponding
interval of $\gamma_b$-s is not narrow

In general, two dissipative factors lead to a decrease of the pitch
angle. The force that provides conservation of adiabatic invariant
$I = 3cp_{\perp}^2/2eB$ in non-uniform magnetic field \citep{landau}
[see also \citep{difus3}]
\begin{equation}\label{g}
G_{\perp} = -\frac{mc^2}{\rho}\gamma_b\psi,\;\;\;\;\;G_{_{\|}} =
\frac{mc^2}{\rho}\gamma_b\psi^2
\end{equation}
and the radiation reaction force \citep{landau}
\begin{equation}\label{f}
F_{\perp} = -\alpha\psi(1 + \gamma_b^2\psi^2),\;\;\;\;\;F_{_{\|}} =
-\alpha\gamma_b^2\psi^2,
\end{equation}
where $\alpha = 2e^2\omega_B^2/(3c^2)$ and $\psi$ is the pitch
angle. Only under the action of these forces the pitch angles tend
to decrease, inevitably killing the synchrotron emission. In reality
the situation is principally different, because the excited
relatively low frequency waves (in our case submillimeter/infrared)
by means of the cyclotron resonance leads to the QLD. Unlike the
dissipative effects of (${\bf F}$, ${\bf G}$), diffusion creates
non-zero distribution function with respect to pitch angles, and
supports the synchrotron emission. It is clear that under favorable
conditions the effect of quasi-linear diffusion may balance the
dissipation, therefore, our objective is to find the distribution of
particles by pitch angles, estimate their mean value and analyze the
corresponding synchrotron emission energy. On the other hand, since
the QLD results from the feedback of the cyclotron modes, apart from
the high energy emission, the system will also be characterized by
the low energy radiation.

In \citep{difus3} we studied the role of the QLD in producing the
X-ray emission by means of ultra-relativistic electrons in AGN
magnetospheric flows. It was shown that the cyclotron resonance
provides emission in a low energy domain - radio band. Unlike the
physical conditions ($|G_{\perp}|\gg |F_{\perp}|$ and
$|G_{_\parallel}|\ll |F_{_\parallel}|$) considered in
\citep{difus3}, in the present paper we examine physically different
regime, $|G_{\perp}|\ll |F_{\perp}|$ and $|G_{_\parallel}|\ll
|F_{_\parallel}|$, which reduces the stationary kinetic equation
governing the quasi-linear diffusion to \citep{malmach}
\begin{equation}\label{kinet}
\frac{\partial}{\partial\psi} \left(\psi
F_{_\perp}f\right)=\frac{1}{mc\gamma_b}\frac{\partial}{\partial\psi}
\left(\psi D_{_{\perp\perp}}\frac{\partial f}{\partial\psi}\right),
\end{equation}
where $f = f(\psi)$ is the distribution function of particles and
\begin{equation}\label{dif}
D_{\perp\perp}\approx \frac{\pi^2
e^2}{m^2c^3}\frac{\delta}{\gamma_b^2}|E_k|^2,
\end{equation}
is the diffusion coefficient. $|E_k|^2$ is the energy density per
unit wavelength, therefore the energy density of the cyclotron waves
are of the order of $|E_k|^2k$. We assume that $\sim 50\%$ of the
resonant plasma energy, $mc^2n_b\gamma_b$, converts to waves
\citep{difus3}, then for an expression of $|E_k|^2$ one obtains
\citep{malmach}
\begin{equation}\label{ek2}
|E_k|^2 = \frac{mc^3n_b\gamma_b} {2\omega}.
\end{equation}
The distribution function obtained from Eq. (\ref{dif}) is given by
\begin{equation}\label{chi} f(\psi) = Ce^{-A\psi^4},
\end{equation}
where
\begin{equation}\label{A}
A\equiv \frac{\alpha mc\gamma_b^3}{4D_{_{\perp\perp}}}, \;\;\;\; C =
const.
\end{equation}
Unlike the work presented in \citep{difus3}, due to the different
regime, $f(\psi)$ behaves as $e^{-A\psi^4}$ instead of
$e^{-A_1\psi^2}$ [see Eq. (11) in \citep{difus3}].

As we see, particles are distributed by the pitch angles, therefore,
the electrons will emit via the synchrotron process without damping.

\begin{figure}
  \resizebox{\hsize}{!}{\includegraphics[angle=0]{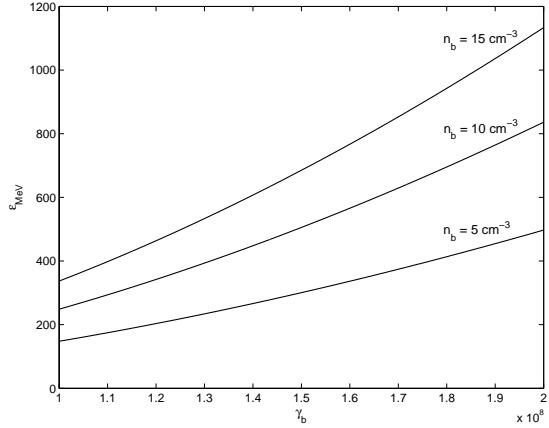}}
  \caption{Behaviour of $\epsilon_{MeV}$ versus $\gamma_b$ for different values of
  $n_b$.
  The set of parameters is $L = 10^{45}erg/s$, $\Omega = 3\times 10^{-5}rad/s$, $\gamma_p =
200$ and $n_b = \{5;10;15\}cm^{-3}$. As is seen from the plots,
relativistic electrons with Lorentz factors $\gamma_b=\{1-2\}\times
10^8$ may provide the high energy radiation in the MeV-GeV domain.}
 \label{fig1}
\end{figure}
\begin{figure}
  \resizebox{\hsize}{!}{\includegraphics[angle=0]{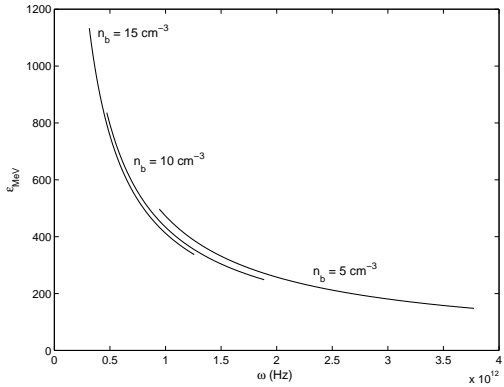}}
  \caption{Behaviour of $\epsilon_{MeV}$ versus $\omega$ for different values of
  $n_b$.
  The set of parameters is $L = 10^{45}erg/s$, $\Omega = 3\times 10^{-5}rad/s$, $\gamma_p =
200$ and $n_b = \{5;10;15\}cm^{-3}$. We see that the high energy
radiation is strongly connected with the submillimeter/low infrared
emission.}
 \label{fig2}
\end{figure}
\begin{figure}
  \resizebox{\hsize}{!}{\includegraphics[angle=0]{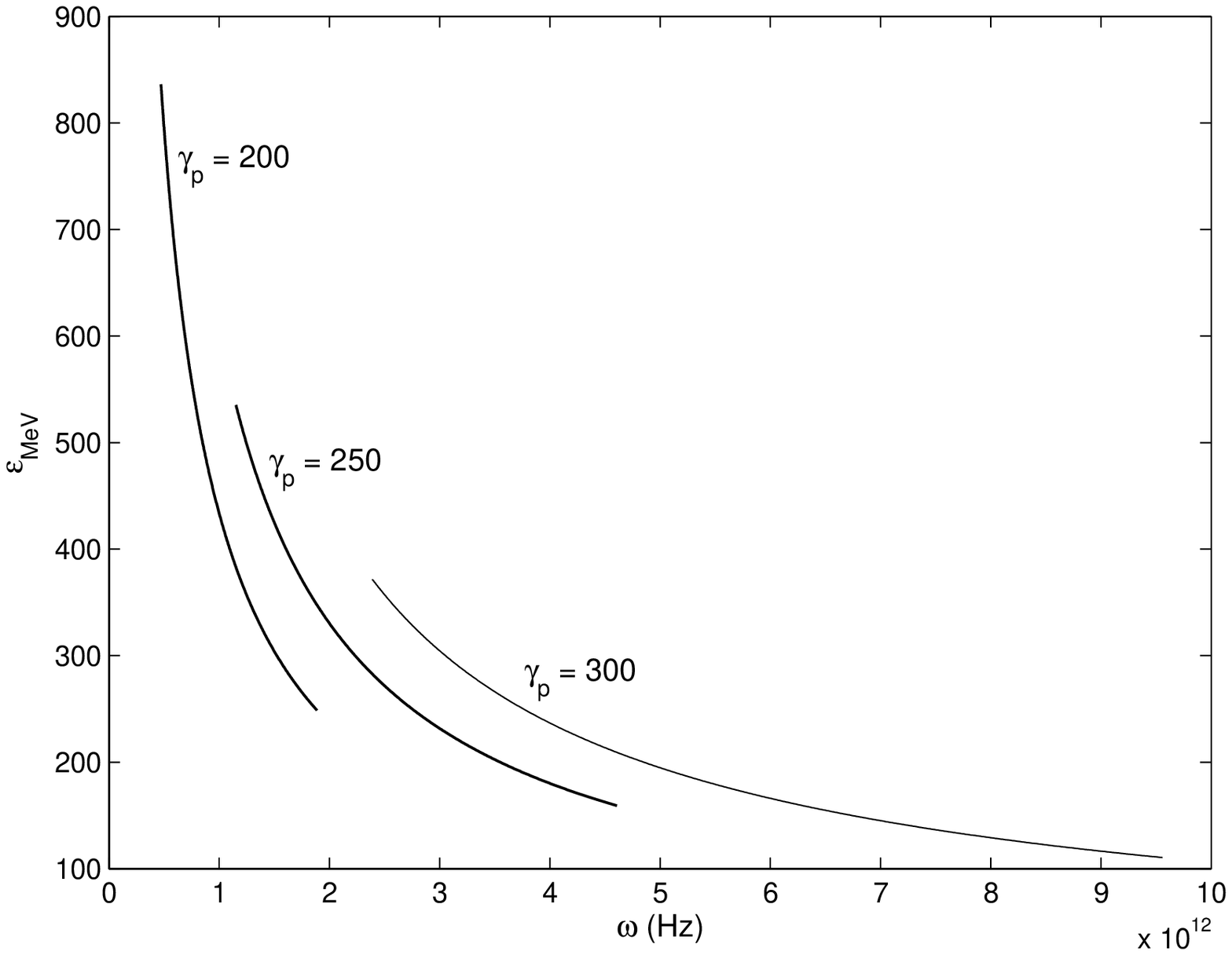}}
  \caption{Behaviour of $\epsilon_{MeV}$ versus $\omega$ for different values of
  $\gamma_p$.
  The set of parameters is $L = 10^{45}erg/s$, $\Omega = 3\times
10^{-5}rad/s$, $\gamma_p = \{200;250;300\}$ and $n_b =
\{5;10;15\}cm^{-3}$. The results show
  the strong connection of $\{100-800\}$MeV radiation with the emission
  in the frequency range $\{0.05-1\}\times 10^{13}Hz$.}
 \label{fig3}
\end{figure}
\section{Discussion}\label{sec:summary}

In this section we apply the model of the QLD to the light cylinder
lengthscales of typical AGNs.

For explaining the high energy radiation, it is strongly believed
that AGN magnetospheres are consist of highly relativistic
electrons. This fact sets another problem - how to accelerate these
particles to such high energies? In general, there are several
mechanisms, which may account for the efficient acceleration of
electrons. Indeed, as is shown in a series of papers, Fermi-type
acceleration process \citep{cw99}, re-acceleration of
electron-positron pairs as a feedback mechanism \citep{ghis} and
centrifugal acceleration \citep{mr94,osm7,ra08,osm10} may provide
very high Lorentz factors of the order of $\gamma_b\sim 10^{5-9}$.
Therefore, in the framework of the paper an existence of such
particles is assumed to be as a given fact. If we suppose an
isotropic distribution of relativistic electrons, one can estimate
the synchrotron cooling timescale \citep{difus3}
\begin{equation}\label{ts}
t_{cool}\approx 5\times
10^{-3}\times\left(\frac{10^2G}{B}\right)^2\times\left(\frac{10^8}{\gamma}\right)s.
\end{equation}
The value of the magnetic induction is given by \citep{difus3}
\begin{equation}
\label{b} B_{lc}\approx
260\times\left(\frac{L}{10^{45}erg/s}\right)^{1/2}\times\left(\frac{\Omega}{3\times
10^{-5}s^{-1}}\right)G,
\end{equation}
where $L$ is the bolometric luminosity of the AGN and $r_{lc} =
c/\Omega$ is the light cylinder radius (a hypothetical zone, where
the linear velocity of rigid rotation exactly equals the speed of
light). $\Omega$ is the magnetic field lines' angular velocity of
rotation, normalized to the value $3\times 10^{-5}s^{-1}$
\citep{belv}. Throughout the paper we assume that the magnetic field
is robust enough to maintain the frozen-in condition in the
magnetosphere of the AGN. Indeed, as one can see, for the typical
magnetospheric parameters, $\gamma_b\sim 10^8$, $n_b\sim 10
cm^{-3}$, the following condition $B_{lc}^2/8\pi>\gamma_b mn_bc^2$
is satisfied, which means that the plasma particles will be forced
to follow the rigidly rotating field lines. During such a motion,
especially on the light cylinder lengthscales, the electrons will
undergo the centrifugal force accelerating them up to very high
Lorentz factors $\sim 10^{8-9}$ \citep{osm7,ra08}

It is clear from Eq. (\ref{ts}) that for a certain class of physical
parameters the synchrotron cooling timescale is of the order of
$5\times 10^{-4}s$. On the other hand, the kinematic timescale of
the system, $t_{kin}\sim r_{lc}/c\sim 3\times 10^4s$ is by many
orders of magnitude bigger than $t_{cool}$, which in turn means that
without the quasi-linear diffusion, particles very soon would stop
emitting in the synchrotron regime, after transiting to their ground
Landau level.

\cite{kmm} showed that the anomalous Doppler effect generates the
cyclotron waves with the frequency \citep{difus3}
$$\omega\approx 6.8\times
10^9\times\left(\frac{\gamma_p}{100}\right)^4\times\left(\frac{10^8}
{\gamma_b}\right)^2\times$$
\begin{equation}
\label{om} \times\left(\frac{B}{100G}\right)^3
\times\left(\frac{10cm^{-3}}{n_b}\right)Hz,
\end{equation}
leading to the process of the quasi-linear diffusion, which, despite
the efficient dissipative factors, creates the pitch angles.

To demonstrate the present model, we consider an AGN with the
bolometric luminosity $L = 10^{45}erg/s$. Let us examine the
following parameters $\Omega = 3\times 10^{-5}rad/s$, $\gamma_b =
10^8$ and $n_b = 10 cm^{-3}$. Since the particles are distributed by
the pitch angles [see Eq. (\ref{chi})], for analyzing the
synchrotron emission it is reasonable to estimate a mean value of
$\psi$
\begin{equation}\label{pitch}
\bar{\psi}
 = \frac{\int_{0}^{\infty}\psi f(\psi)d\psi}{\int_{0}^{\infty}f(\psi)d\psi}
\approx \frac{0.5}{\sqrt[4]{A}}.
\end{equation}
Then one can show from Eq. (\ref{pitch}) that for the aforementioned
parameters the pitch angle is of the order of $8\times 10^{-3}rad$,
therefore, relativistic electrons will inevitably emit photons with
energies \citep{Lightman}
\begin{equation}
\label{eps} \epsilon_{eV}\approx 1.2\times 10^{-8}B\gamma^2\sin\psi.
\end{equation}
After substituting the value of $\overline{\psi}$ in Eq.
(\ref{eps}), we see that the synchrotron emission generates
radiation in the MeV domain.

The quasi-linear diffusion works if the cyclotron modes are excited,
therefore, it is essential to estimate the timescale of the
corresponding instability ($t_{ins}$) and compare it with the
kinematic timescale of the system. According to the work of
\cite{kmm} the growth rate of the instability is given by
\begin{equation}\label{inc1}
\Gamma = \pi \frac{\omega_b^2}{\omega\gamma_p} \;\;\;\ if \;\;\;\
\frac{1}{2}\frac{u_x^2}{c^2}\ll\delta
\end{equation}
and
\begin{equation}\label{inc2}
\Gamma = \pi
\frac{\omega_b^2}{2\omega\gamma_p}\frac{u_x^2}{\delta\cdot c^2}
\;\;\;\ if \;\;\;\ \frac{1}{2}\frac{u_x^2}{c^2}\gg\delta,
\end{equation}
where $\omega_b\equiv\sqrt{4\pi n_b e^2/m}$ is the plasma frequency
of beam electrons. It is easy to show that for $n_b = 10cm^{-3}$,
$\gamma_p = 200$ (see Fig. \ref{fig1}), $V_{_{\parallel}}\sim c$ and
$\rho\sim R_g$, one obtains $u_x^2/(2c^2)\ll 1$, implying that the
increment of the instability is given by Eq. (\ref{inc1}). The
cyclotron resonance makes sense if $t_{ins}/t_{kin}<1$, then, by
taking into account the definition of the kinematic timescale,
$r_{lc}/c$ and the instability timescale, $1/\Gamma$, one can show
that the aforementioned condition reduces to
\begin{equation}
\label{cond}3.5\times10^{-3}\times\left(\frac{\gamma_p}{100}\right)^5\times\left(\frac{10^8}
{\gamma_b}\right)^2\times\left(\frac{10cm^{-3}}{n_b}\right)^2<1.
\end{equation}
As is clear from Eq. (\ref{cond}), the condition is very sensitive
to the Lorentz factor of the plasma components, and for relatively
higher values of $\gamma_p$ the condition will violate. The upper
limit of $\gamma_p$, when the condition is still valid is of the
order of $300$ for $\gamma_b\sim 10^8$ and $n_b\sim 10cm^{-3}$.

For studying the efficiency of the QLD, we examine the following set
of the parameters $L = 10^{45}erg/s$, $\Omega = 3\times
10^{-5}rad/s$, $\gamma_p = 200$ and $n_b = \{5;10;15\}cm^{-3}$ and
the results are demonstrated in Fig.\ref{fig1} where we show the
behaviour of $\epsilon_{_{MeV}}$ versus $\gamma_b$. From the plots
is clear that $\epsilon_{_{MeV}}$ is a continuously increasing
function of the beam Lorentz factor, which is a natural result of
the fact that more energetic particles produce more energetic
photons. The behaviour of $\epsilon_{_{MeV}}$ versus $n_b$ is
different, more dense beam electrons produce photons with lower
energies. This can be seen from Eqs. (\ref{A},\ref{pitch}):
$\bar{\psi}\sim\sqrt[4]{D_{_{\perp\perp}}}$, which by combining with
$D_{_{\perp\perp}}\sim n_b^3$ [see Eqs. (\ref{dif},\ref{ek2})]
confirms the dependence $\epsilon_{_{MeV}}(n_b)$. According to the
results demonstrated in the figure, relativistic electrons with
Lorentz factors $\gamma_b=\{1-2\}\times 10^8$ may provide the high
energy radiation in the MeV-GeV domain.

Since the generation of the synchrotron emission strongly depends on
the cyclotron waves, we also investigate the behaviour of
$\epsilon_{_{MeV}}$ versus $\omega$. Figure \ref{fig2} shows the
function $\epsilon_{_{MeV}}(\omega)$ for several values of $n_b$.
The set of parameters is the same as in the previous figure. As is
clear from the plots, the $\{200-1200\}$MeV radiation is strongly
connected with the submillimeter ($\sim [0.3-3]\times 10^{12}Hz$)
and low infrared ($\sim [3-3.8]\times 10^{12}Hz$) emission.

Another important parameter, the physical system depends on, is the
Lorentz factor of the plasma component. Therefore, it is reasonable
to demonstrate the function $\epsilon_{_{MeV}}(\omega)$ for
different values of $\gamma_p$. These results are shown in Fig.
\ref{fig3}, where the set of parameters is $L = 10^{45}erg/s$,
$\Omega = 3\times 10^{-5}rad/s$, $\gamma_p = \{200;250;300\}$ and
$n_b = \{5;10;15\}cm^{-3}$. As we see, the excited infrared domain
extends up to $\sim 10^{13}Hz$, which is strongly connected with
high energy emission ($100MeV$). As is clear from the figure, higher
values of $\gamma_p$ correspond to lower synchrotron energies.
Indeed, by taking into account the relation
$\bar{\psi}\sim\sqrt[4]{D_{_{\perp\perp}}}$ combined with Eqs.
(\ref{dif},\ref{ek2}) one can see that $\bar{\psi}\sim
1/\gamma_p^2$.

As is clear from the results, the quasi-linear diffusion together
with the cyclotron instability may guarantee production of high
energy radiation in the MeV-GeV domains strongly connected with
submillimeter/infrared emission. The major difference in results
from our previous work is that in \citep{difus3} we studied physical
conditions leading to excitation of $X$-rays connected with the
relatively low frequency radio band (KHz-MHz), whereas in the
present paper both energies (produced as by synchrotron as by
cyclotron mechanisms respectively) are much higher. This
investigation sets another problem: since the QLD is a feasible
mechanism providing the aforementioned high energies, one of the
important next steps could be testing of MeV-GeV AGNs exhibiting an
efficient submillimeter/infrared radiation and see if the strong
correlation is observationally evident. This in turn, could be a
certain test for estimating the AGN magnetospheric parameters, such
as the density and the Lorentz factors of plasma component and beam
electrons. A particular future objective is to investigate
theoretically the radiative signatures of both high and low energy
emissions respectively. This work we are going to perform sooner or
later.

\section{Summary}\label{sec:summary}

The main aspects of the present work can be summarized as follows:
\begin{enumerate}

      \item Mechanisms producing strongly connected high and low energy
      radiation was studied by taking into account the QLD in the AGN
      magnetospheres. Considering a physical regime different from
      that of \citep{difus3}, we investigate the efficiency of the QLD
      in a region close to the light cylinder surface.

      \item For the considered physical parameters, it has been
      shown that the cyclotron instability appears for relatively
      low frequency range, producing radiation in the submillimeter/infrared
      domains. On the other hand, despite the short cooling timescales,
      the effect of diffusion on particles recreates the pitch
      angles and produces the high energy radiation in the MeV-GeV
      bands.

      \item The problem was studied versus three major
      magnetospheric parameters: the beam and the plasma
      component's Lorentz factors and the beam electrons' density.
      It was shown that the photon energy, $\epsilon_{_{MeV}}$, is a continuously
      increasing function of the beam Lorentz factor and the beam
      density. Contrary to this, by increasing the plasma component
      Lorentz factor, the corresponding photon energy decreases.

      \end{enumerate}

\section*{Acknowledgments}
The research was supported by the Georgian National Science
Foundation grants GNSF/ST06/4-096 and GNSF/ST06/4-193.


%
\end{document}